# A NOTE ON THE IMPLEMENTATION OF THE HYPERELASTIC KILIAN MODEL IN THE ABAQUS

*Aleksander FRANUS*

## 1. Introduction

The chapter addresses a study of well-known Kilian [6] and Gent [3] hyperelastic material models. A key feature of these models is that they describe so called strain locking. The constraint results directly from a structure of the stored energy function. It is worth to mention that in the theory of small deformations, constitutive relations for 'locking materials' [2] are defined by analogy to the plasticity theory which is not the case here.

In order to illustrate some basic properties of the considered incompressible [5], strain locking material models, we investigate three types of homogeneous deformations [4,7,8]. Solutions to the problems are derived in terms of the prescribed displacements. Moreover, basic traction problems are discussed. In particular, a form of instability is shown in a block subjected to biaxial tension.

The main goal is to verify the implementation of the Kilian material model in the Abaqus/Standard [1]. The software does not provide any straightforward way to treat properly user's material models exhibiting a strain locking. One way around is to implement user-defined element (UEL) including hybrid formulation based on the Lagrange function [9] with desired constraints. The way is not consistence with the constraint results directly from a structure of the stored energy function and only mentioned here. However, we show that available in the materials library the Kilian model describes properly locking constraint only in a simple test performed on one finite element. In the case of the UHYPER procedure [1] such constraint is not fulfilled. Results of three tests based on the built-in model and the procedure are presented. The documentation of the Abaqus does not contain any information concerning implementation of the constraint regarding the model.

## 2. Hyperelasticity

We begin by setting up basic definitions for large strain nonlinear elasticity. A material is called hyperelastic, cf. [5], if there exists a stored-energy function such that $W : \Im \to [0, +\infty], \Im = \{\mathbf{F} \in L^2(\Omega, \mathbb{R}) : \det \mathbf{F} > 0\}$ with

$$\mathbf{S} = \mathrm{D}_\mathbf{F} W(\mathbf{F}), \tag{2.1}$$

where $\mathbf{S}$ is the first Piola-Kirchhoff stress tensor and $\mathbf{F}$ defines a 'deformation gradient' as $\mathbf{F} = \mathrm{D}_\mathbf{X} \chi(\mathbf{X})$. The function $\chi : \Omega \ni \mathbf{X} \to \mathbf{x} \in \mathbb{R}^3$ defines an actual configuration of the body with respect to an initial one. In order to avoid interpenetration of matter, it is necessary to put some restrictions on $W(\mathbf{F})$, see [7]:

$$W(\mathbf{F}) \to +\infty \quad \text{as } J \to 0^+, \quad W(\mathbf{F}) \to +\infty \quad \text{as } \|\mathbf{F}\| \to +\infty. \tag{2.2}$$

A stored-energy function is isotropic if

$$W(\mathbf{F}) = \tilde{W}(\mathbf{C}) = \tilde{W}(\mathbf{B}) = \breve{W}(I_1, I_2, I_3), \tag{2.3}$$

where $I_1, I_2$ and $I_3$ are the invariants of each of the left and right Cauchy-Green deformation tensors: $\mathbf{B} = \mathbf{F}\mathbf{F}^T, \mathbf{C} = \mathbf{F}^T \mathbf{F}$. Since $\mathbf{S}$ is non-symmetric, it is convenient to introduce the second Piola-Kirchhoff stress tensor

$$\begin{aligned}\mathbf{T} &= \mathrm{D}_\mathbf{E} \hat{W}(\mathbf{E}) = 2\mathrm{D}_\mathbf{C} \tilde{W}(\mathbf{C}) = 2(\gamma_1 \mathbf{I} + \gamma_2 \mathbf{C} + \gamma_3 \mathbf{C}^2),\\ \gamma_1 &= \frac{\partial \breve{W}}{\partial I_1} + \frac{\partial \breve{W}}{\partial I_2} I_1 + \frac{\partial \breve{W}}{\partial I_3} I_2, \quad \gamma_2 = -\left(\frac{\partial \breve{W}}{\partial I_2} + \frac{\partial \breve{W}}{\partial I_3} I_1\right), \quad \gamma_3 = \frac{\partial \breve{W}}{\partial I_3}.\end{aligned} \tag{2.4}$$

The relationship between the Cauchy stress tensor $\boldsymbol{\sigma}$, the first and second Piola-Kirchhoff ones, namely $\boldsymbol{\tau} = J\boldsymbol{\sigma} = \mathbf{S}\mathbf{F}^T = \mathbf{F}\mathbf{T}\mathbf{F}^T$, cf. [5], implies a constitutive relation in the spatial description such that

$$\begin{aligned}\boldsymbol{\tau} &= 2\mathbf{B}\mathrm{D}_\mathbf{B} \tilde{W}(\mathbf{B}) = 2(\beta_1 \mathbf{I} + \beta_2 \mathbf{B} + \beta_3 \mathbf{B}^2),\\ \beta_1 &= \frac{\partial \breve{W}}{\partial I_3} I_3, \quad \beta_2 = \frac{\partial \breve{W}}{\partial I_1} + \frac{\partial \breve{W}}{\partial I_2} I_1, \quad \beta_3 = -\frac{\partial \breve{W}}{\partial I_2}.\end{aligned} \tag{2.5}$$

A material with internal constrains $J = \det \mathbf{F} = 1$ is called an incompressible one [5]. In a context of a boundary value problem it is convenient to define a Lagrange function such that

$$L = \bar{W}(\bar{\mathbf{C}}) - p(\det \mathbf{C} - 1), \quad \bar{\mathbf{C}} = \bar{\mathbf{F}}^T \bar{\mathbf{F}}, \bar{\mathbf{F}} = J^{-\frac{1}{3}} \mathbf{F} \tag{2.6}$$

The second Piola-Kirchhoff stress tensor has the form

$$\mathbf{T} = 2\mathrm{D}_\mathbf{C} \bar{W}(\bar{\mathbf{C}}) - Jp\mathbf{C}^{-1} \tag{2.7}$$

A Lagrange multiplier $p$ can be interpreted as hydrostatic pressure. The formulation with the modified deformation gradient $\bar{\mathbf{F}}$ with the volume change eliminated may be easily extend to slightly compressible material models.

The other typical constraint in a material is inextensibility [9]. To enforce that a material does not extend in the direction $\mathbf{m}$, the Lagrange multiplier formulation takes the form

$$L = \breve{W}(\mathbf{C}) - q(\mathrm{tr}(\mathbf{MC}) - 1), \quad \mathbf{M} = \mathbf{m} \otimes \mathbf{m}, \qquad (2.8)$$

where the multiplier may physically represent a fiber stress related to the constraint $\mathrm{tr}(\mathbf{MC}) = 1$. The way of enforcing a strain locking is not consistent with the constraint that results directly from a structure of the stored energy function and only mentioned here.

## 3. Constitutive models

One of the simplest hyperelastic model describing strain locking is known due to Gent [2]. The stored energy function is stated as

$$W_G = -\frac{\mu_0}{2} a \ln\left(1 - \frac{\bar{I}_1 - 3}{a}\right) \qquad (3.1)$$

with $\mu_0$ as initial shear modulus and $a$ bounds $\bar{I}_1$ such that $\bar{I}_1 < 3 + a$. When $a \to \infty$, the neo-Hookean model is recovered as

$$\lim_{a \to \infty} W_G = \frac{\mu_0}{2}(\bar{I}_1 - 3). \qquad (3.2)$$

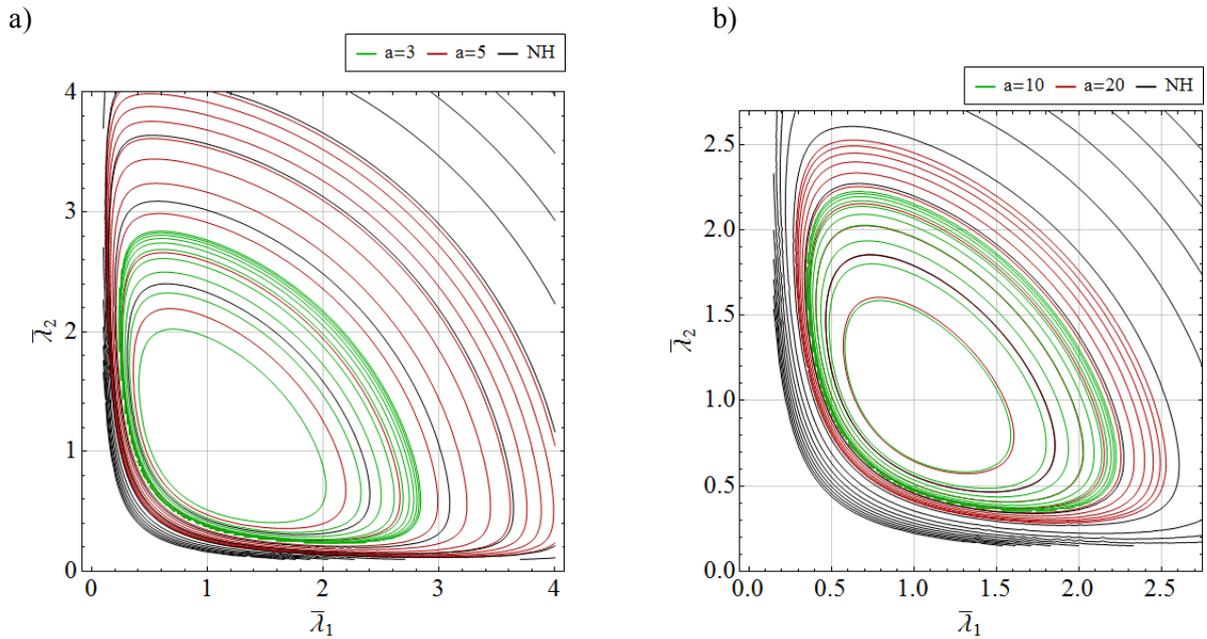

**Figure 3.1.** Contour plots of the stored energy function a) the Kilian model b) the Gent model.

The hyperelastic Kilian model [6], also known as Van der Waals due to an analogy in the thermodynamic interpretation of the equations of state for rubber and gas, is described by the function

$$W_K = -\mu_0 \left\{ (a^2-3)[\ln(1-\eta)+\eta] - \frac{2}{3}\alpha\left(\frac{\tilde{I}-3}{2}\right)^{3/2} \right\}, \eta = \sqrt{(\tilde{I}-3)/(a^2-3)} \quad (3.3)$$

together with $\tilde{I} = (1-f)\bar{I}_1 + f\bar{I}_2$. Response of the material is bounded such that $\tilde{I} < a^2$. Similarly to the Van der Waals equation for a real gas, $\alpha$ stands for the global interaction parameter.

Again, in the case of $a \to \infty$, the neo-Hookean is recovered. From now on we assume that $\alpha = f = 0$ which means that (3.3) is a function of the first invariant $\bar{I}_1$ only.

## 4. Homogeneous deformations

### 4.1. Prescribed displacement

In order to illustrate some basic properties of the considered incompressible material models, we consider three types of homogeneous deformations. First of them is a simple shear deformation with the deformation gradient

$$\mathbf{F} = \mathbf{I} + \gamma \mathbf{b}_1 \otimes \mathbf{b}_2, \quad (4.1)$$

where $\gamma$ is the amount of shear, see Fig. 4.1. In the case of the Gent model, non-zero components of the Cauchy stress tensor in considered coordinates system take forms

$$\sigma_{12} = \mu_0 \frac{a\gamma}{a-\gamma^2}, \sigma_{11} = \frac{2}{3}\mu_0 \frac{a\gamma^2}{a-\gamma^2}, \sigma_{22} = \sigma_{33} = -\frac{1}{3}\mu_0 \frac{a\gamma^2}{a-\gamma^2}. \quad (4.2)$$

We see that for this type of deformation the constraint $\bar{I}_1 < 3+a$ leads to $\gamma^2 < a$. Value $\sqrt{a}$ has a clear interpretation, e.g. a limit of the amount of shear.

In the case of the Kilian model, with asymptotes described by equation $\gamma^2 = a^2 - 3$, we have

$$\sigma_{12} = \mu_0 \frac{\gamma}{1-b}, \sigma_{11} = \frac{1}{3}\mu_0 \frac{2\gamma^2}{1-b}, \sigma_{22} = \sigma_{33} = -\frac{1}{3}\mu_0 \frac{\gamma^2}{(1-b)}, \quad b = \sqrt{\frac{\gamma^2}{a^2-3}}. \quad (4.3)$$

Another type of homogeneous deformation is uniaxial stretch, which in a case of incompressible material model is described by

$$\mathbf{F} = \bar{\lambda}_1 \mathbf{b}_1 \otimes \mathbf{b}_1 + \frac{1}{\sqrt{\bar{\lambda}_1}}(\mathbf{b}_2 \otimes \mathbf{b}_2 + \mathbf{b}_3 \otimes \mathbf{b}_3). \quad (4.4)$$

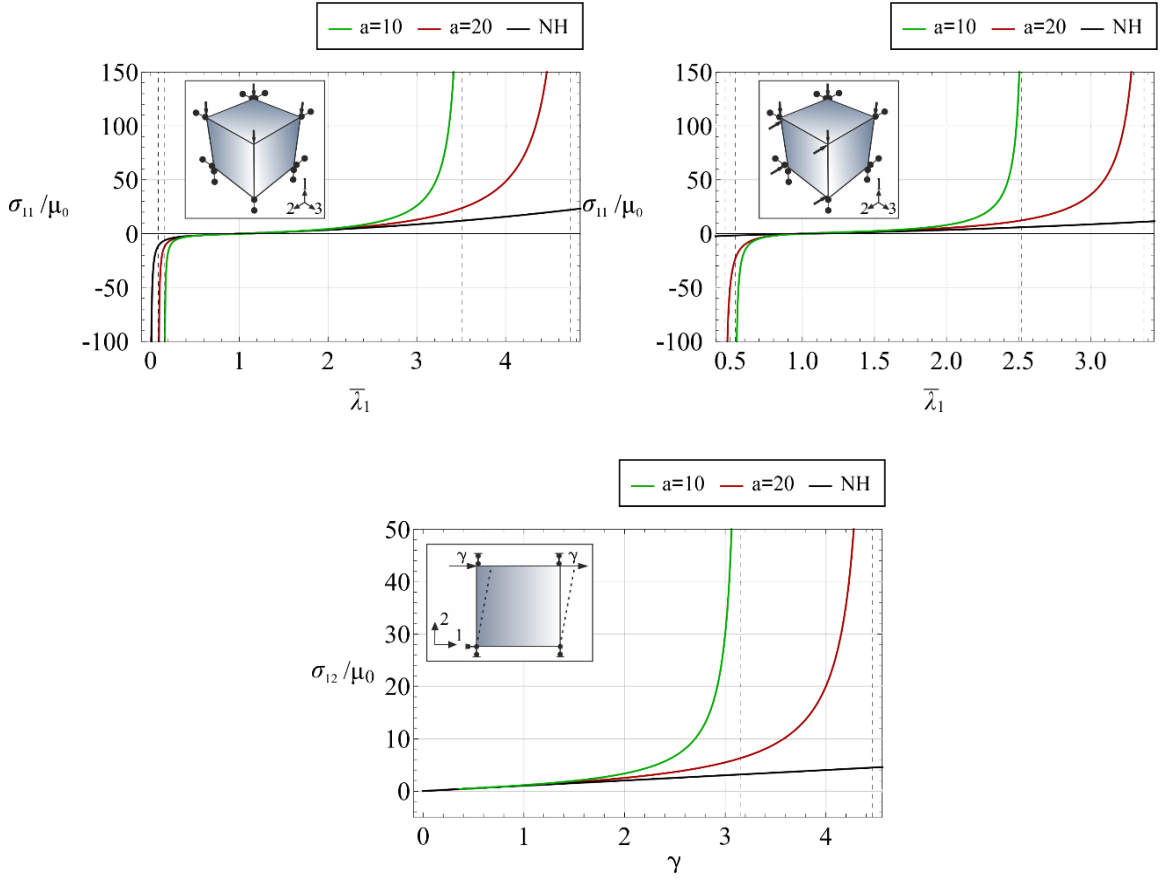

**Figure 4.1.** Selected components of the Cauchy stress tensor in the case of the Gent and neo-Hookean (NH) models: 1) uniaxial deformation, 2) biaxial deformation, 3) simple shear deformation.

Only one component of the Cauchy stress tensor is non-zero. The stored energy function of the Gent model leads to

$$\sigma_{11} = \mu_0 \frac{a(\bar{\lambda}_1^2 - 1)}{-\bar{\lambda}_1^3 - (3+a)\bar{\lambda}_1 + 2} \tag{4.5}$$

with asymptotes as real roots of one-parameter cubic equation $\bar{\lambda}_1^3 + (3+a)\bar{\lambda}_1 - 2 = 0$, cf. Fig. 4.1. Similarly, we derive the formulas and constraint in the case of the Kilian model

$$\sigma_{11} = \mu_0 \frac{1 - \bar{\lambda}_1^3}{\bar{\lambda}_1(b-1)}, \quad b = \sqrt{\frac{(\bar{\lambda}_1 - 1)^2(\bar{\lambda}_1 + 2)}{(a^2 - 3)\bar{\lambda}_1}}, \quad \frac{2}{\bar{\lambda}_1} + \bar{\lambda}_1^2 < a^2. \tag{4.6}$$

Finally, we consider a biaxial stretch with equal principal stretches $\bar{\lambda}_1 = \bar{\lambda}_2$. The deformation gradient reads

$$\mathbf{F} = \bar{\lambda}_1(\mathbf{b}_1 \otimes \mathbf{b}_1 + \mathbf{b}_2 \otimes \mathbf{b}_2) + \frac{1}{\bar{\lambda}_1^2}\mathbf{b}_3 \otimes \mathbf{b}_3. \tag{4.7}$$

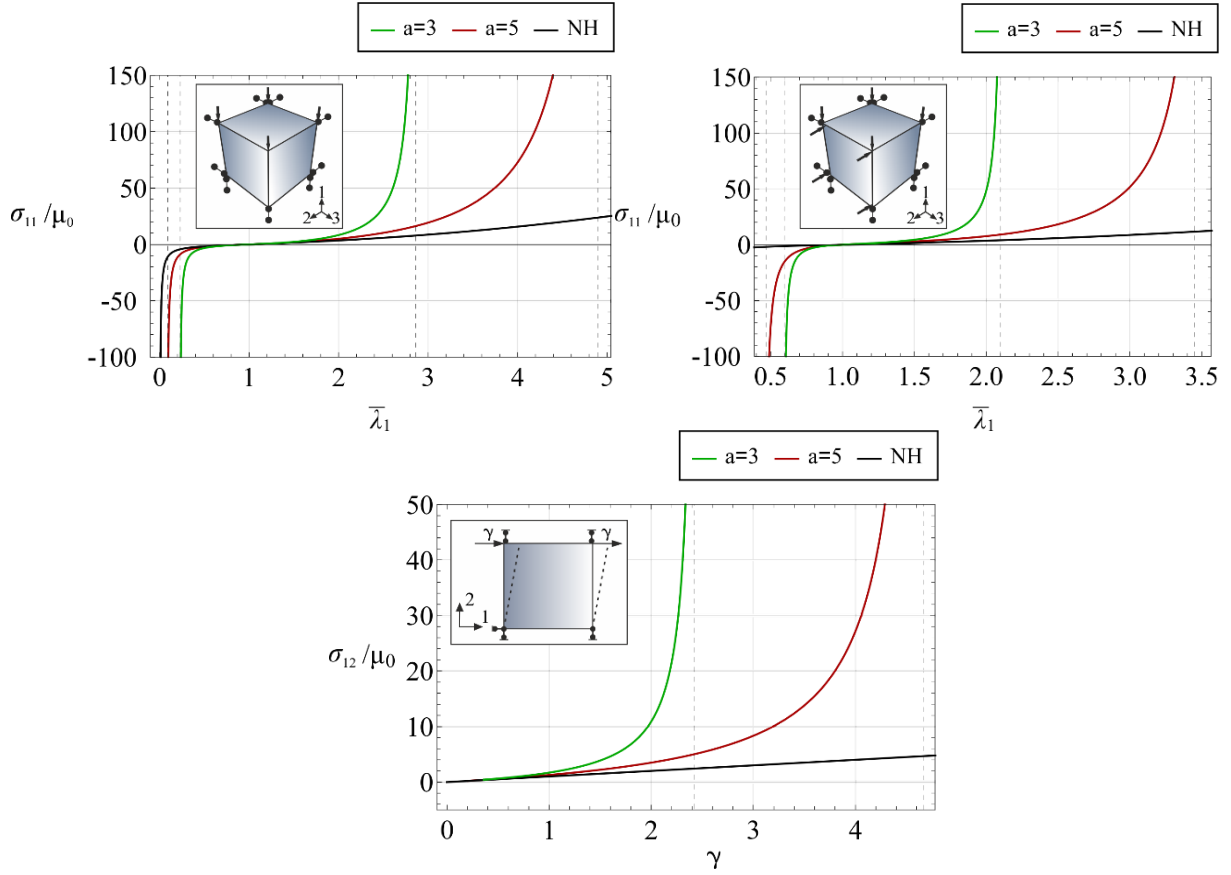

**Figure 4.2.** Selected components of the Cauchy stress tensor in the case of the Kilian and neo-Hookean (NH) models: 1) uniaxial deformation, 2) biaxial deformation, 3) simple shear deformation.

The non-zero components of the Cauchy stress tensor and the constraint in the of Gent model are given by

$$\sigma_{11} = \sigma_{22} = \mu_0 \frac{a\left(\bar{\lambda}_1^6 - 1\right)}{-2\bar{\lambda}_1^6 + (3+a)\bar{\lambda}_1^4 - 1}, \quad \frac{1}{\bar{\lambda}_1^4} + 2\bar{\lambda}_1^2 < 3 + a. \qquad (4.8)$$

The Kilian models in this case yields

$$\sigma_{11} = \sigma_{22} = \mu_0 \frac{1 - \bar{\lambda}_1^6}{\bar{\lambda}_1^4 (b-1)}, \quad b = \sqrt{\frac{2\bar{\lambda}_1^2 + \frac{1}{\bar{\lambda}_1^4} - 3}{a^2 - 3}}, \quad \frac{1}{\bar{\lambda}_1^4} + 2\bar{\lambda}_1^2 < a^2. \qquad (4.9)$$

We see that except for the simple shear deformation, the constraints derived on the basis of the considered models typically lead to asymptote in subset $\bar{\lambda}_1 \in (0,1)$ which is an undesirable feature, because of the fact that physically locking occurs in materials only in a case of extension.

### 4.2. Prescribed traction

Basic properties of the considered incompressible material models are also illustrated in a case of non-zero traction boundary conditions. We consider a cube subjected to the prescribed traction, which is normal to face '1' and '2' with the same magnitude $\tilde{S}$ in a reference con-figuration (dead load). For simplicity we also assume $\bar{\lambda}_3 = 1$. The equilibrium equations and incompressibility condition, cf. [7], lead to

$$\tilde{S}(\bar{\lambda}_1 - \bar{\lambda}_2) - \bar{\lambda}_1 \frac{\partial W}{\partial \bar{\lambda}_1} - \bar{\lambda}_2 \frac{\partial W}{\partial \bar{\lambda}_2} = 0 \tag{4.10}$$

together with $\bar{\lambda}_1 \bar{\lambda}_2 = 1$. A trivial solution occurs with $\bar{\lambda}_1 = \bar{\lambda}_2$. The other one, namely

$$\tilde{S} = \frac{1}{(\bar{\lambda}_1 - \bar{\lambda}_2)} \left( \bar{\lambda}_1 \frac{\partial W}{\partial \bar{\lambda}_1} + \bar{\lambda}_2 \frac{\partial W}{\partial \bar{\lambda}_2} \right), \quad \bar{\lambda}_1 = \frac{1}{\bar{\lambda}_2}, \tag{4.11}$$

describes a non-trivial equilibrium path. The bifurcation occurs at $\tilde{S} = 2\mu_0$ in the case of considered stored energy functions. We see that the Kilian material model is much stiffer for chosen values of the locking parameter, see Fig. 4.3.

The second considered type of homogeneous deformation with the prescribed traction is shown in Fig. 4.4. The traction is normal to face '1' with a magnitude $\tilde{S}$ and the plane strain deformation is assumed. In this case only one type of equilibrium path is derived

$$\sigma_{11} = \tilde{S}\lambda_1 = \bar{\lambda}_1 \frac{\partial W}{\partial \bar{\lambda}_1} - \bar{\lambda}_2 \frac{\partial W}{\partial \bar{\lambda}_2} \tag{4.12}$$

together with $\bar{\lambda}_1 \bar{\lambda}_2 = 1$. However, it does not mean that these are only solutions to the problems as the unknown in the equations is $\bar{\lambda}_1$.

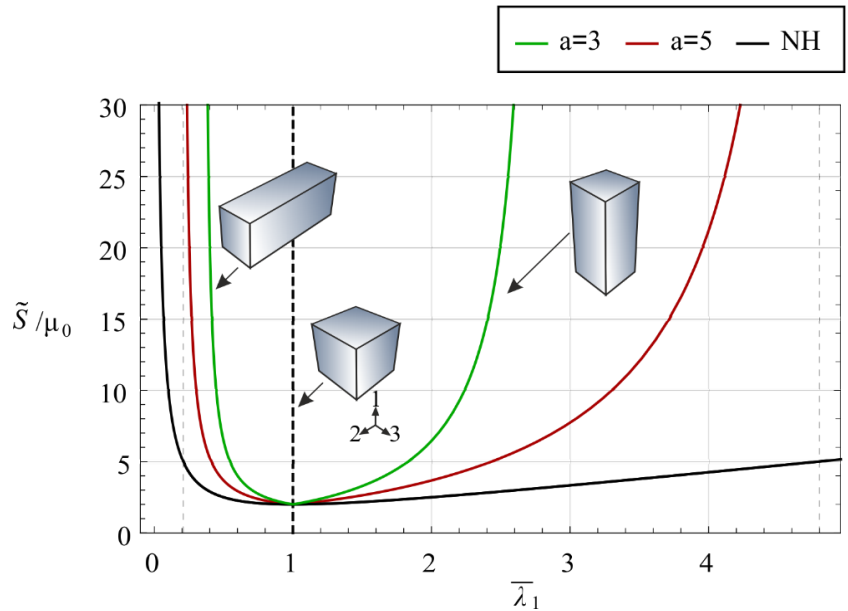

**Figure 4.3.** Boundary traction-principal stretch diagrams for a cube subjected to the prescribed traction – the Kilian and neo-Hookean (NH) models.

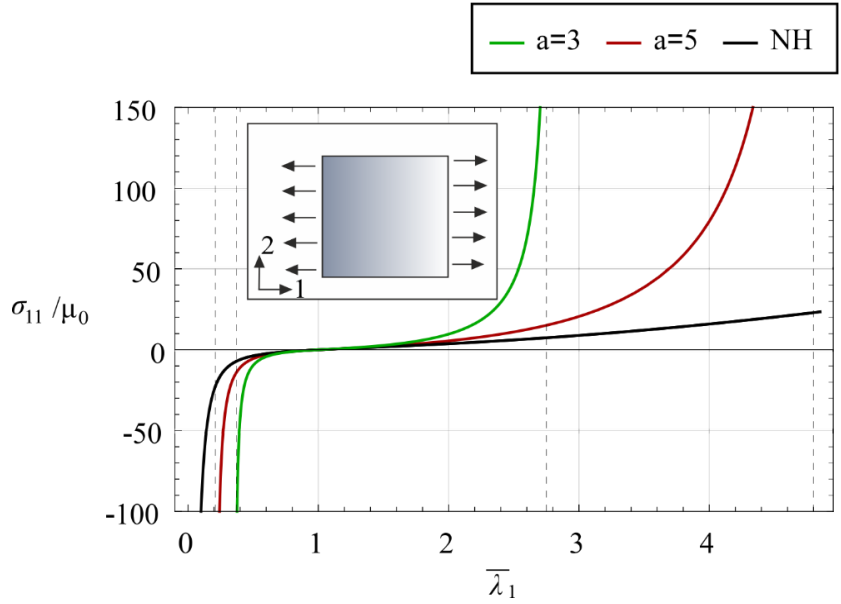

**Figure 4.4.** The Cauchy stress-principal stretch diagrams for a block subjected to the prescribed traction – the Kilian and neo-Hookean (NH) models.

## 5. The Abaqus/Standard

### 5.1. User subroutine UHYPER

One way to define user hyperelastic material model is to implement UHYPER subroutine [1]. One needs to write Fortran code which includes a stored-energy function and its derivatives with respect to the invariants

$$\bar{I}_1 = \mathrm{tr}\bar{\mathbf{B}}, \quad \bar{I}_2 = \frac{1}{2}\left(\bar{I}_1^2 - \mathrm{tr}\bar{\mathbf{B}}^2\right), \quad J = \det \mathbf{F}, \tag{5.1}$$

where $\bar{\mathbf{B}} = \bar{\mathbf{F}}\bar{\mathbf{F}}^T, \bar{\mathbf{F}} = J^{-1/3}\mathbf{F}$. Generally, the ABAQUS/Standard software adopts an approach based on decoupling constitutive relation into three components: isochoric, volumetric and coupling term. In the case of an incompressible model, it is sufficient to define the isochoric part such that

$$\boldsymbol{\tau} = -p\mathbf{I} + \mathbf{s} = -p\mathbf{I} + 2\left(\frac{\partial U}{\partial \bar{I}_1} + \bar{I}_1 \frac{\partial U}{\partial \bar{I}_2}\right)\mathrm{dev}\bar{\mathbf{B}} - 2\frac{\partial U}{\partial \bar{I}_2}\mathrm{dev}\left(\bar{\mathbf{B}}^2\right). \tag{5.2}$$

The Zaremba-Jaumann objective stress rate leads to

$$\overset{\circ}{\mathbf{s}} = \mathfrak{C}_d \cdot \mathrm{dev}\mathbf{D}, \tag{5.3}$$

where in a case of stored energy function of the form $W = U(\bar{I}_1)$ the elasticity tensor in the spatial description [1] reads

$$\mathfrak{C}_d = 4\frac{\partial^2 U}{\partial \bar{I}_1^2}\bar{\mathbf{B}} \otimes \bar{\mathbf{B}} - \frac{4}{3}\left[\frac{\partial U}{\partial \bar{I}_1} + \bar{I}_1 \frac{\partial^2 U}{\partial \bar{I}_1^2}\right](\mathbf{I} \otimes \bar{\mathbf{B}} + \bar{\mathbf{B}} \otimes \mathbf{I}) + 2\frac{\partial U}{\partial \bar{I}_1}(\mathbf{I} \lozenge \bar{\mathbf{B}} + \bar{\mathbf{B}} \lozenge \mathbf{I}). \tag{5.4}$$

The derivatives of the stored energy function of the Kilian model take forms

$$\frac{\partial U}{\partial \bar{I}_1} = \frac{\mu_0}{2\left(1 - \sqrt{\frac{\bar{I}_1 - 3}{a^2 - 3}}\right)}, \quad \frac{\partial^2 U}{\partial \bar{I}_1^2} = \frac{\mu_0 \sqrt{\frac{\bar{I}_1 - 3}{a^2 - 3}}}{4\left(\sqrt{\frac{\bar{I}_1 - 3}{a^2 - 3}} - 1\right)^2 (\bar{I}_1 - 3)}. \quad (5.5)$$

It is worth to notice that the second order derivative is singular in a case of the natural state, i.e. $\bar{I}_1 - 3 = 0$.

## 5.2. Verification of the Kilian model in the Abaqus/Standard

The Abaqus software does not provide any straightforward way to treat properly user's material models exhibiting strain locking. One way is to implement user-defined element (UEL) including hybrid formulation based on the Lagrange function (2.8). However, we show that available in the materials the Kilian model describes properly locking constraint only in a simple test performed on one finite element while the UHYPER procedure does not. Documentation of the Abaqus [1] does not contain any information concerning implementation of the constraint regarding the Kilian model.

In order to verify the material model in Abaqus/Standard two simple numerical tests are carried out based on the built-in Kilian model and UHYPER procedure. First of them is a pure shear problem with a traction boundary conditions using the 4-node bilinear plane strain quadrilateral, hybrid element (CPE4H), cf. Fig. 5.1. The material parameters are assumed as $a = 5, \alpha = f = 0$. Second one concerns a similar problem, but with the prescribed displacement, cf. Fig. 5.2.

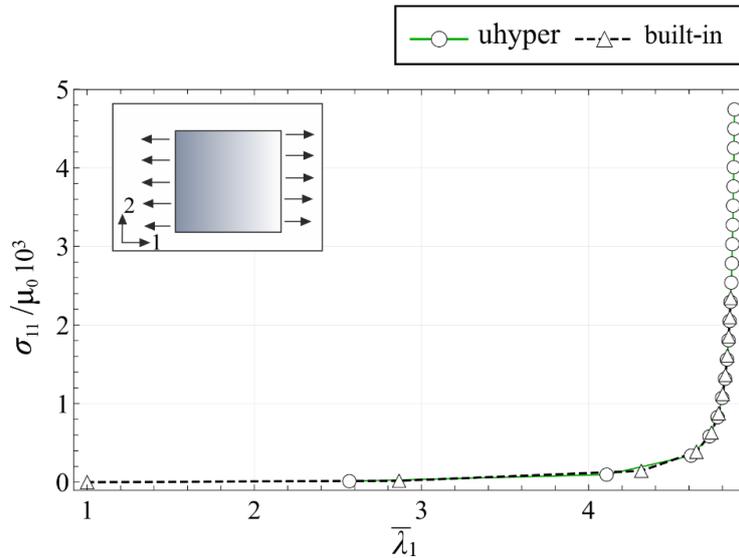

**Figure 5.1.** The Cauchy stress-principal stretch diagrams for a block subjected to the prescribed traction – the Kilian model.

Fig. 5.1 and 5.2 present the Cauchy stress $\sigma_{11}$ as a function of principal deformation $\bar{\lambda}_1$ corresponding to the traction and displacement boundary conditions respectively. Firstly, it is worth to notice that a solution in the prescribed traction problem is obtained in a greater range in terms of stress values using UHYPER procedure. Clearly both solutions approach to asymptote $\bar{\lambda}_1 = 4.96$, which means that the strain constraint is satisfied. These results are the same as the ones derive analytically, cf. subsection 4.2.

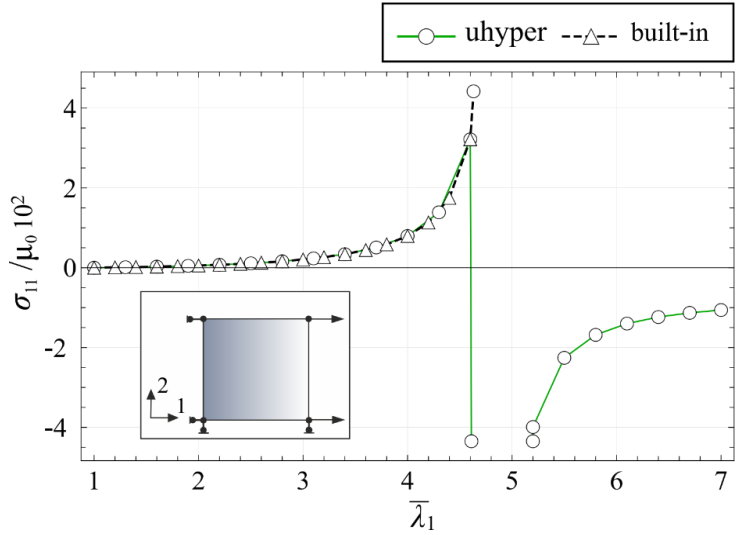

**Figure 5.2.** The Cauchy stress-principal stretch diagrams for a block subjected to the prescribed displacement – the Kilian model.

More interesting situation takes place in the prescribed displacement problem. In the case of UHYPER procedure values of the deformation are not bounded properly, see Fig. 5.2. Evidently solver skips the asymptote and continues a procedure. It happens using the Gent model as well. It is not the case when it comes to the built-in Kilian model. The solution approaches to the same asymptote as in the traction problem. This indicates existing additional numerical procedure implemented in the Abaqus's solver which keeps a solution bounded.

Better insight in the procedure is given by results of test with nonhomogeneous deformations. The model consists of three elements with hybrid, plane strain formulation (CPE4H), see Fig. 5.3. Displacement boundary conditions are prescribed: on one face $u_1 = u_2 = 0$ and on the opposite face $u_2$ is controlled. The material parameters are assumed as $a = 3, \alpha = f = 0$.

Presented results of logarithmic strain in direction 1 show that in the case of the built-in model, algorithm converges only till the locking strain occurs in one element. In terms of stretch in considered direction we have $\exp(1.023) = 2.78$ which is approximately equal to locking stretch in homogeneous deformation. This indicates that corresponding locking constraint is checked based on strain on each element separately. The approach is not acceptable, because it should consider the constraint for globally deformed body. In the case of the UHYPER procedure clearly the constraint is not hold. Finial configurations of the mesh are shown in Fig. 5.3.

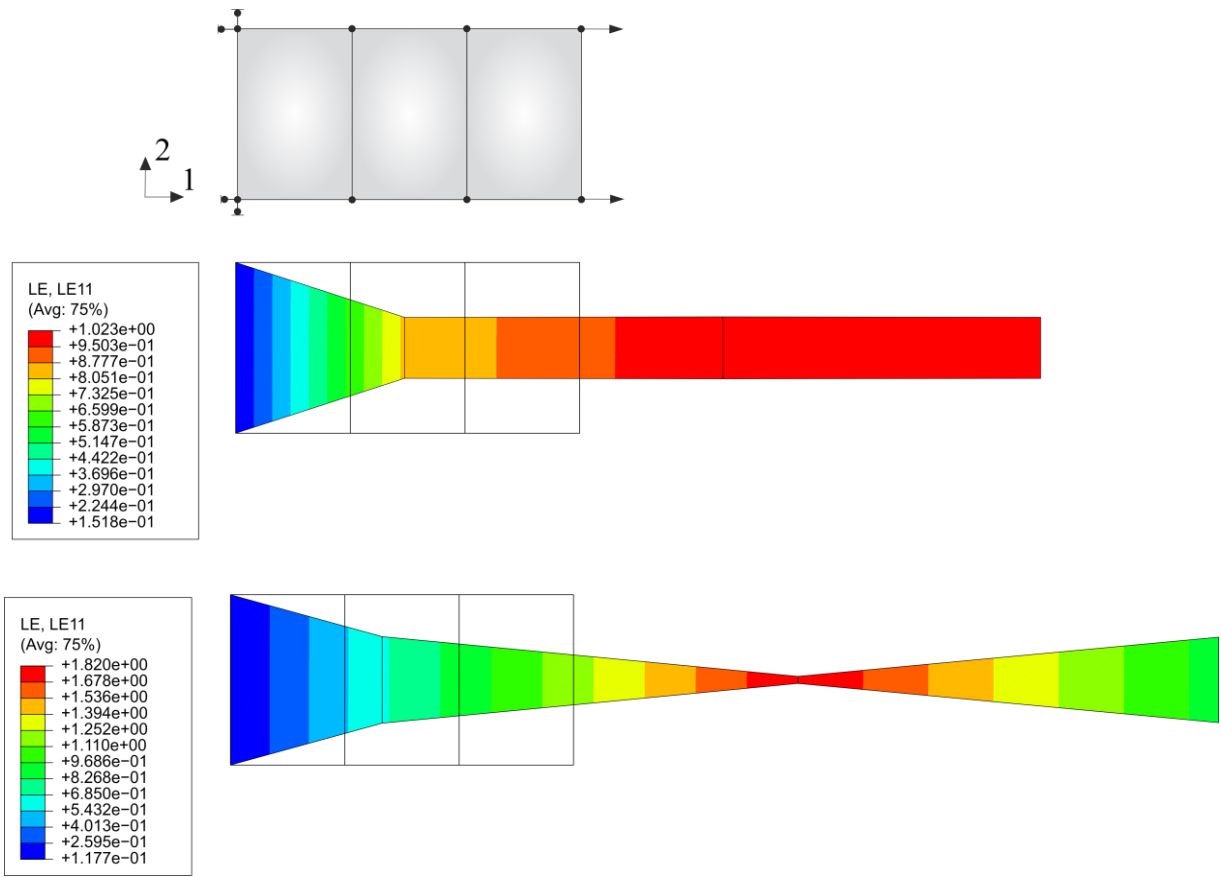

**Figure 5.3.**  Contour plots of logarithmic strain in direction 1 – comparison of results obtained based on the built-in Kilian model (a) and the UHYPER procedure (b).

## 6. Conclusions

The Abaqus software does not provide any straightforward way to treat properly user's material models exhibiting a strain locking. However, it is shown that available in the materials library the Kilian model describes properly locking constraint only in a simple test performed on one finite element. In the case of the presented finite element model consisted of three elements, the Abaqus's algorithm converges only till the locking strain occurs in one of them. The result shows that corresponding locking constraint is checked based on strain on each element separately. The approach is not acceptable, because it should consider the constraint for globally deformed body. To provide this, a constitutive relation should be formulated by analogy to the plasticity theory which is not the case here. Nevertheless, the documentation of the Abaqus does not contain any information concerning implementation of the constraint regarding the model. In the case of the UHYPER procedure such constraint is not fulfilled in any of the presented cases.